\documentclass[aip,apl,reprint]{revtex4-1}

\usepackage{graphicx}
\usepackage{dcolumn}
\usepackage{amsmath}
\usepackage[normalem]{ulem}

\makeatletter
\def\btt#1{\texttt{\@backslashchar#1}}
\DeclareRobustCommand\bblash{\btt{\@backslashchar}}
\makeatother

\usepackage{color}

\begin{document}

\preprint{HEP/123-qed}

%%%%%%%%%%%%%%%%%%%%%%%%%%%%%%%%%%%%%%%%%%%%%%%%%%%%%%%%%%%%%%%%%%%%%%%%%%%%%%%

\title[Short Title]{Circular dichroism in a three-dimensional semiconductor chiral photonic crystal}

\author{S. Takahashi}
\affiliation{
Institute of Nano Quantum Information Electronics, University of Tokyo, 4-6-1 Komaba, Meguro-ku, Tokyo 153-8505, Japan
}

\author{T. Tajiri}
\affiliation{
Institute of Industrial Science, University of Tokyo, 4-6-1 Komaba, Meguro-ku, Tokyo 153-8505, Japan
}

\author{Y. Ota}
\affiliation{
Institute of Nano Quantum Information Electronics, University of Tokyo, 4-6-1 Komaba, Meguro-ku, Tokyo 153-8505, Japan
}

\author{J. Tatebayashi}
\affiliation{
Institute of Nano Quantum Information Electronics, University of Tokyo, 4-6-1 Komaba, Meguro-ku, Tokyo 153-8505, Japan
}

\author{S. Iwamoto}
\affiliation{
Institute of Nano Quantum Information Electronics, University of Tokyo, 4-6-1 Komaba, Meguro-ku, Tokyo 153-8505, Japan
}
\affiliation{
Institute of Industrial Science, University of Tokyo, 4-6-1 Komaba, Meguro-ku, Tokyo 153-8505, Japan
}

\author{Y. Arakawa}
\affiliation{
Institute of Nano Quantum Information Electronics, University of Tokyo, 4-6-1 Komaba, Meguro-ku, Tokyo 153-8505, Japan
}
\affiliation{
Institute of Industrial Science, University of Tokyo, 4-6-1 Komaba, Meguro-ku, Tokyo 153-8505, Japan
}

%%%%%%%%%%%%%%%%%%%%%%%%%%%%%%%%%%%%%%%%%%%%%%%%%%%%%%%%%%%%%%%%%%%%%%%%%%%%%%%

\date{\today}

%%%%%%%%%%%%%%%%%%%%%%%%%%%%%%%%%%%%%%%%%%%%%%%%%%%%%%%%%%%%%%%%%%%%%%%%%%%%%%%

\begin{abstract}
Circular dichroism covering the telecommunication band is experimentally demonstrated in a semiconductor-based three-dimensional chiral photonic crystal (PhC).
We design a rotationally-stacked woodpile PhC structure where neighboring layers are rotated by $60^{\circ}$ and three layers construct a single helical unit.
The mirror-asymmetric PhC made from GaAs with sub-micron periodicity is fabricated by a micro-manipulation technique.
Due to the large contrast of refractive indices between GaAs and air, the experimentally obtained circular dichroism extends over a wide wavelength range, with the transmittance of right-handed circularly polarized incident light being 85\% and that of left-handed light being 15\% at a wavelength of 1.3 $\mu$m.
The obtained results show good agreement with numerical simulations.
\end{abstract}

%%%%%%%%%%%%%%%%%%%%%%%%%%%%%%%%%%%%%%%%%%%%%%%%%%%%%%%%%%%%%%%%%%%%%%%%%%%%%%%

%\pacs{Valid PACS appear here}

\maketitle

%%%%%%%%%%%%%%%%%%%%%%%%%%%%%%%%%%%%%%%%%%%%%%%%%%%%%%%%%%%%%%%%%%%%%%%%%%%%%%%

The eigenpolarization of light passing through a purely chiral structure\cite{Gansel}, along its chiral or helical axis, is circular polarization (CP).
Analogous to achiral structures, where the eigenpolarization is linear polarization, periodic chiral structures show Bragg reflection for circularly polarized light which has the same chirality as the structures.
This phenomena leads to a difference in transmission between right-handed CP (RCP) and left-handed CP (LCP), resulting in circular dichorism.
CP-based filters and cavities using circular dichroism can be conceived from linear-polarization-based achiral photonic crystals (PhCs), and circularly polarized laser oscillation has been achieved in self-assembled cholesteric liquid crystals\cite{review,Coles}.
In addition to the reports of circular Bragg reflection in liquid crystals, it has also been reported in many materials, such as self-assembled plasmonic nanoparticles\cite{Kuzyk}, artificially patterned or deposited metamaterials\cite{Gansel,Zhao,Gibbs,Decker}, polymers\cite{Thiel1,Thiel2,Gu}, plasmonic oligomers\cite{Hentschel}, and complexes\cite{Hrudey}.
However, circular Bragg reflection has been scarcely reported in semiconductors.

%%%%%%%%%%%%%%%%%%%%%%%%%%%%%%%%%%%%%%%%%%%%%%%%%%%%%%%%%%%%%%%%%%%%%%%%%%%%%%%

Semiconductor materials are regarded as lossless dielectrics with large refractive indices for light with frequency below the bandgap energy.
Since the bandwidth of circular dichroism observed in chiral structures is determined by the anisotropy of the refractive indices in the plane perpendicular to the helical axis\cite{review}, neatly designed chiral structures based on semiconductors can show broadband circular dichroism with little loss.
In addition, semiconductor-based chiral structures can be applied to highly efficient lasers emitting CP light and CP-based spin-photon interfaces\cite{Greve,Gao} for future optical circuits or quantum information processing.
These applications may also benefit from the fact that efficient photon emitters can be easily introduced in semiconductor systems.
Although the fabrication of semiconductor-based three-dimensional (3D) structures is still challenging, recently a micro-manipulation technique\cite{Aoki1,Aoki2} has been used to realize GaAs-based 3D PhCs with high quality\cite{Aniwat}.
Using this technique, we have fabricated a GaAs-based chiral PhC and demonstrated large optical rotation\cite{Takahashi}, which is another effect of optical activity.
In this letter we demonstrate a 3D chiral PhC made of GaAs, and broadband circular dichroism in the telecommunication wavelength region.

%%%%%%%%%%%%%%%%%%%%%%%%%%%%%%%%%%%%%%%%%%%%%%%%%%%%%%%%%%%%%%%%%%%%%%%%%%%%%%%

The studied 3D chiral PhC, called a rotationally-stacked woodpile structure\cite{Takahashi}, is schematically shown in Fig. 1(a).
The structure is composed of a stack of patterned thin layers each of thickness 225 nm.
Each layer of the stack is comprised of periodic rods with 160 nm width and $a$ = 500 nm period.
The rods are made of GaAs (with refractive index $n = 3.4$), and they are surrounded by air.
These layers are stacked one by one with a $60^{\circ}$ in-plane rotation in order that three layers construct a single helical unit corresponding to a space group with $3_1$ screw operation.
In this study, we stacked 16 layers and constructed a 3D chiral PhC with the total thickness of 3.6 $\mu$m.
We define the $x$ ($y$) axis as those orthogonal (parallel) to the rods in the bottom layer, as shown in Fig. 1(a).

%%%%%%%%%%%%%%%%%%%%%%%%%%%%%%%%%%%%%%%%%%%%%%%%%%%%%%%%%%%%%%%%%%%%%%%%%%%%%%%

\begin{figure}[t]
\includegraphics[width=1.0\linewidth]{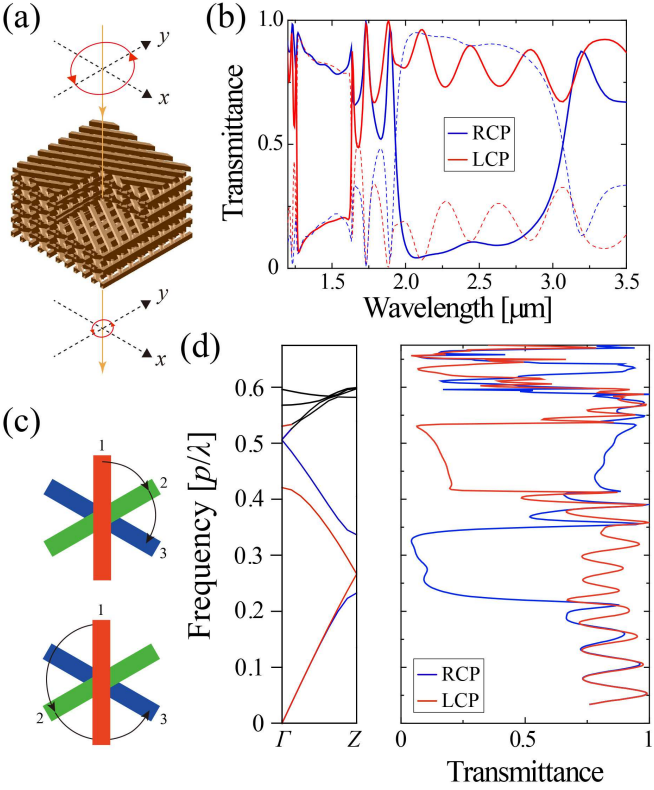}
\caption{\label{fig:calculation}
(a) Schematic of the studied 3D chiral PhC.
The top (16th) layer is removed and a part of the structure is cut for clarity.
(b) Calculated transmittance spectra for RCP (blue) and LCP (red) incidence.
Two wavelength regions of circular dichroism are found with opposite transmittance relation between the two orthogonal CPs.
The reflectance spectra are also plotted as dotted lines.
(c) Schematic top view of the chiral PhC.
The studied structure has both chiralities with different helical unit cells.
(d) Band structure along the direction perpendicular to the stacked layers (left) and the corresponding transmittance spectra for each CP incidence (right).
The right figure is the same numerical result as (b) but plotted by the normalized frequency.
The circular dichroism regions are consistent through the two numerical results.
}
\end{figure}

%%%%%%%%%%%%%%%%%%%%%%%%%%%%%%%%%%%%%%%%%%%%%%%%%%%%%%%%%%%%%%%%%%%%%%%%%%%%%%%

We first performed numerical simulations for such a chiral structure to investigate circular dichroism using a finite-difference time domain (FDTD) method.
Periodic boundary conditions are imposed in both the $x$ and $y$ directions, and in the stacking direction, perfectly matched layers are attached at 10 $\mu$m far from the center of the structure.
The simulated incident light is a pulsed plane wave with either RCP or LCP, and incident normally on the top of the structure as shown in Fig. 1(a).
The pulse duration is set to be a single cycle for the central wavelength (1.75 $\mu$m).
The time-dependent electric fields of the transmitted light are recorded for each CP incidence, and we analyze the spectral response using a Fourier transform.
The transmittance is calculated by the ratio between two numerical results with and without the structure.

%%%%%%%%%%%%%%%%%%%%%%%%%%%%%%%%%%%%%%%%%%%%%%%%%%%%%%%%%%%%%%%%%%%%%%%%%%%%%%%

Figure 1(b) shows the calculated transmittance spectra for RCP and LCP incident light.
In two wavelength regions, 1.2 $\mu$m $<$ $\lambda$ $<$ 1.6 $\mu$m and 2.0 $\mu$m $<$ $\lambda$ $<$ 3.0 $\mu$m, the transmittance is largely different between the RCP and LCP incidence, indicating circular dichroism.
Note that the circular dichroism is caused by the differential reflectance for different CPs rather than the differential absorption as shown with dotted lines in Fig. 1(b).
Such broadbands of circular dichroism are caused by the chiral PhC made of GaAs which has a large refractive index contrast with air.
However, the relation between RCP and LCP is opposite in the two circular dichroism regions.
This dual-band circular dichroism is caused by the coexistence of two helical structures in the chiral PhC, a left-handed helix with $120^{\circ}$ rotation per single layer and a right-handed helix with $60^{\circ}$ rotation per single layer as shown in Fig. 1(c)\cite{Hodgkinson,Lee1}.
In other words, two different unit cells composed of three or six layers coexist in the structure.
Since one of the CPs with the same handedness as the structural chirality is reflected by circular Bragg reflection, LCP is reflected around $\lambda$ = 1.4 $\mu$m which is almost half of the RCP reflection wavelength, reflecting the number of layers constructing each unit cell.
In general, such dual-band circular dichroism appears in discrete helical structures constructing single helical unit with $N$ layers.
Circular Bragg reflection occurs for one of the CPs at a wavelength corresponding to $N$ layers as well as for the other CP at a wavelength corresponding to $360^{\circ}/(180^{\circ}-360^{\circ}/N)$ layers.
In this study, the dual-band circular dichroism appears in the two wavelength ranges for $N$ = 3.
Note that the region between the two opposite circular dichroism regions can be used to show large optical rotation as in our previous report\cite{Takahashi}.

%%%%%%%%%%%%%%%%%%%%%%%%%%%%%%%%%%%%%%%%%%%%%%%%%%%%%%%%%%%%%%%%%%%%%%%%%%%%%%%

The photonic band structure of the chiral PhC in the direction along the helical axis is also calculated by a plane wave expansion method and is shown as the left figure in Fig. 1(d).
In this figure, the dispersion relation of each optical mode is plotted with the frequency normalized by the helical pitch $p$ = 0.675 $\mu$m and with the wave number in the first Brillouin zone from the origin $\it {\Gamma}$ to $Z$ (which is the edge of the zone corresponding to the direction along the helical axis in the real space).
The optical modes in the blue (red) colored bands have strong RCP (LCP)\cite{Lee2}, and polarization gaps appear around $p/\lambda$ = 0.25 and 0.5.
The right panel in Fig. 1(d) shows the same numerical result as Fig. 1(b) but is plotted by the normalized frequency.
Upon comparing the two figures, it is clear that the circular dichroism regions show good agreement between the two different numerical methods.
Note that since the FDTD calculations are performed for the chiral PhC with 16 layers, those results do not show the complete suppression of each CP in the circular dichroism regions.

%%%%%%%%%%%%%%%%%%%%%%%%%%%%%%%%%%%%%%%%%%%%%%%%%%%%%%%%%%%%%%%%%%%%%%%%%%%%%%%

\begin{figure}[t]
\includegraphics[width=1.0\linewidth]{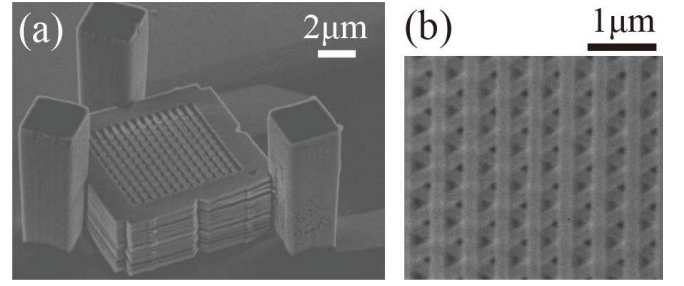}
\caption{\label{fig:sample}
(a) Scanning electron micrograph (SEM) image of the fabricated 3D chiral PhC.
16 layers are stacked using three posts as a guide.
(b) SEM image of the periodic rods.
The crossing points arranged in a triangle lattice are aligned along the helical axes.
}
\end{figure}

%%%%%%%%%%%%%%%%%%%%%%%%%%%%%%%%%%%%%%%%%%%%%%%%%%%%%%%%%%%%%%%%%%%%%%%%%%%%%%%

To investigate the circular dichroism experimentally, we fabricated a 3D GaAs-based chiral PhC.
Plates were patterned with the same dimensions as in the numerical simulation by electron beam lithography and dry and wet etching.
Then, the processed plates were stacked on a GaAs substrate with guide posts using micro-manipulation\cite{Aoki1,Aoki2,Aniwat} as shown in Fig. 2.
For this sample, we applied circularly polarized light along the helical axis at room temperature and measured its transmitted power.
First, we used a super-continuum laser as a light source and polarized the laser into each CP using broadband wave plates.
The polarization-synthesized laser beam was then focused on the sample by a $20\times$ objective lens with a relatively small numerical aperture (0.45) in order that the depth of focus $\sim$ 5 $\mu$m is larger than the total thickness of the structure, enabling us to irradiate approximate plane waves on the sample.
The intensity spectrum of the transmitted light was measured by a monochromator with an InGaAs photodetector.
Figure 3 shows the observed transmission spectra for both RCP and LCP incident light.
The transmitted power for the RCP incidence exceeds that for the LCP incidence in the wavelength range of 1.3 $\mu$m $<$ $\lambda$ $<$ 1.6 $\mu$m, while there is little difference for $\lambda$ $<$ 1.1 $\mu$m.
This is the direct observation of broadband circular dichroism covering the telecommunication band, as caused by the chiral PhC.

%%%%%%%%%%%%%%%%%%%%%%%%%%%%%%%%%%%%%%%%%%%%%%%%%%%%%%%%%%%%%%%%%%%%%%%%%%%%%%%

\begin{figure}[t]
\includegraphics[width=0.75\linewidth]{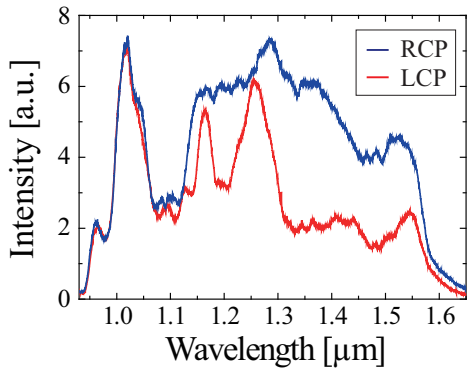}
\caption{\label{fig:experiment_sc}
Transmission spectra through the chiral PhC measured by a super-continuum light source and a monochromator with an InGaAs photodetector.
A large transmission difference for RCP and LCP appears around the telecommunication band, indicating broadband circular dichroism, while there is little difference below 1.1 $\mu$m.
}
\end{figure}

%%%%%%%%%%%%%%%%%%%%%%%%%%%%%%%%%%%%%%%%%%%%%%%%%%%%%%%%%%%%%%%%%%%%%%%%%%%%%%%

%%%%%%%%%%%%%%%%%%%%%%%%%%%%%%%%%%%%%%%%%%%%%%%%%%%%%%%%%%%%%%%%%%%%%%%%%%%%%%%

\begin{figure}[t]
\includegraphics[width=1.0\linewidth]{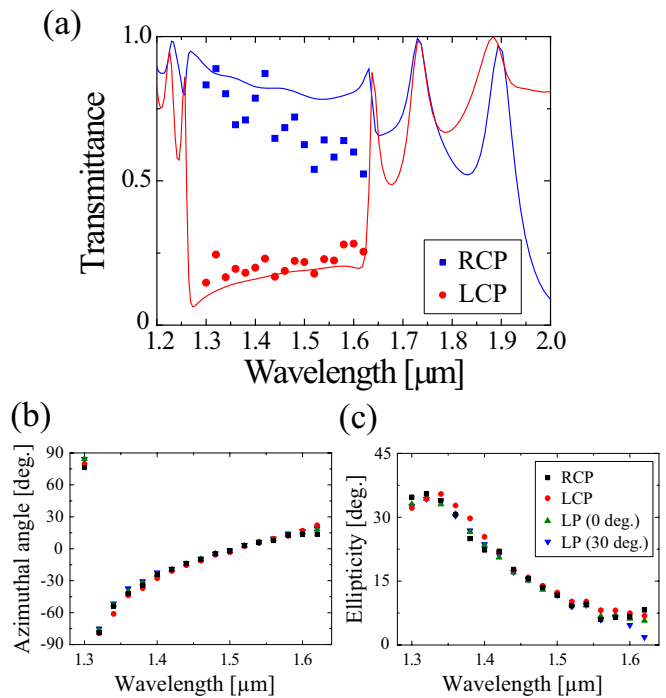}
\caption{\label{fig:experiment_mono}
(a) Observed transmittance spectra for RCP and LCP plotted as blue and red dots, respectively.
Circular dichroism appears throughout the measured wavelength region.
The simulated results in Fig. 1(b) are also plotted as two solid lines.
(b), (c) Plots of azimuthal angle and ellipticity of the polarization in the transmitted light for RCP, LCP, linear polarization (LP) along the $x$ axis, and linear polarization with $30^{\circ}$ azimuthal rotation angle measured from the $x$ axis.
Every polarization of incidence gives almost same results, indicating eigenpolarization in the chiral structure.
}
\end{figure}

%%%%%%%%%%%%%%%%%%%%%%%%%%%%%%%%%%%%%%%%%%%%%%%%%%%%%%%%%%%%%%%%%%%%%%%%%%%%%%%

In order to measure this broadband circular dichroism precisely, we then adopted a monochromatic wavelength-tunable laser with a range of 1.3 $\mu$m $<$ $\lambda$ $<$ 1.62 $\mu$m as a light source, and measured the power and polarization of the transmitted light by using a quarter wave plate, a linear polarizer, and a power meter.
Figure 4(a) shows the observed transmittance spectra for the incident beams with RCP and LCP, together with the numerical results from Fig. 1(b).
Note that the experimental transmittance is obtained by measuring the ratio between the transmitted power through both the 3D PhC and the substrate with that through the substrate only.
The transmitted power for each CP incidence is very different throughout the measured wavelength region.
The transmittance of the RCP is 85\% while that of the LCP is 15\% at a wavelength of 1.3 $\mu$m; differing in transmission by a factor of 6.
These results agree well with the numerical calculations.
The slight difference between the experimental and numerical results is probably due to the finite numerical aperture of the objective lens.
Imperfection in the fabrication and the experimental setup could be other possible reasons.

%%%%%%%%%%%%%%%%%%%%%%%%%%%%%%%%%%%%%%%%%%%%%%%%%%%%%%%%%%%%%%%%%%%%%%%%%%%%%%%

The azimuthal angle and ellipticity of the transmitted light at various input beam polarizations and wavelengths are shown in Fig. 4(b) and (c).
From these figures, we can see that the polarization of the transmitted light is independent of the incident light (we applied RCP, LCP, and two different linear polarizations).
This result indicates that the transmitted elliptic polarization is one of the eigenpolarizations in the studied 3D chiral PhC and the light with the other orthogonal eigenpolarization is reflected.

%%%%%%%%%%%%%%%%%%%%%%%%%%%%%%%%%%%%%%%%%%%%%%%%%%%%%%%%%%%%%%%%%%%%%%%%%%%%%%%

In summary, we have investigated circular dichroism in a rotationally-stacked woodpile PhC structure made from GaAs.
Numerical calculations show dual-band circular dichroism which is inherent to discrete helical structures.
We fabricated the 3D chiral PhC with GaAs and measured the transmitted power through the structure for RCP and LCP incidence.
Owing to the large contrast of refractive indices between GaAs and air, we found a broadband circular dichroism with 6 times transmittance difference between the two CP incidences, showing good agreement with the numerical simulation.
We expect that the obtained circular dichroism can be tuned to any wavelength region by scaling the dimension of the chiral PhC.
This result will pave the way to the realization of CP-based devices in semiconductors, negative phase velocity of light in chiral structures without metal\cite{book}, or even to biological and medical field\cite{Agranat}.

%%%%%%%%%%%%%%%%%%%%%%%%%%%%%%%%%%%%%%%%%%%%%%%%%%%%%%%%%%%%%%%%%%%%%%%%%%%%%%%

This work was supported by the Project for Developing Innovation Systems of MEXT and JSPS through its FIRST Program.

%%%%%%%%%%%%%%%%%%%%%%%%%%%%%%%%%%%%%%%%%%%%%%%%%%%%%%%%%%%%%%%%%%%%%%%%%%%%%%%

%%%%%%%%%%%%%%%%%%%%%%%%%%%%%%%%%%%%%%%%%%%%%%%%%%%%%%%%%%%%%%%%%%%%%%%%%%%%%%%

\end{document}